\documentclass[onecolumn,amsmath,amssymb,nofootinbib,12pt]{article}
\usepackage{jheppub}

\usepackage{graphicx}
\usepackage{dcolumn}
\usepackage{bm,psfrag}

\usepackage{subfigure}
\usepackage{float}
\usepackage{tensor}

\newcommand{\ben}{\begin{equation}}
\newcommand{\een}{\end{equation}}
\newcommand{\be}{\begin{equation}}
\newcommand{\ee}{\end{equation}}
\newcommand{\bea}{\begin{eqnarray}}
\newcommand{\eea}{\end{eqnarray}}
\newcommand{\ba}{\begin{eqnarray}}
\newcommand{\ea}{\end{eqnarray}}

\newcommand{\beq}{\begin{equation}}
\newcommand{\eeq}{\end{equation}}
\newcommand{\beqa}{\begin{eqnarray}}
\newcommand{\eeqa}{\end{eqnarray}}
\newcommand{\beqar}{\begin{eqnarray*}}
\newcommand{\eeqar}{\end{eqnarray*}}

\def\t6 {T_\mt{D6}}

\newcommand{\mt}[1]{\textrm{\tiny #1}}

\newcommand{\vk}{{\vec{k}}}

\def\cale         {{\cal E}}

\def\ee           {{\rm e}}

\def\sqr#1#2{{\vcenter{\vbox{\hrule height.#2pt
 \hbox{\vrule width.#2pt height#1pt \kern#1pt
 \vrule width.#2pt}\hrule height.#2pt}}}}

\def\ee{\cale}

\def\aa1{\phi}
\def\cc1{\psi}

\newcommand{\dt}{\delta t}

\begin{document}

\preprint{arXiv:1902.02945 [hep-th]}

\title{Complexity and scaling in quantum quench in $1+1$ dimensional fermionic field theories} 

\author{Sinong Liu$^1$}
\affiliation{$^1$\,Department of Physics and Astronomy, University of Kentucky,\\ 
\vphantom{k}\ \ Lexington, KY 40506, USA}

\emailAdd{sinong.liu@uky.edu}

\date{\today}

\abstract{We consider the scaling behavior of circuit complexity under quantum quench in an a relativistic fermion field theory on a one dimensional spatial lattice. This is done by finding an exactly solvable quench protocol which asymptotes to massive phases at early and late times and crosses a critical point in between. We find a variety of scaling behavior as a function of the quench rate, starting with a saturation for quenches at the lattice scale, a "fast quench scaling" at intermediate rate and a Kibble Zurek scaling at slow rates. }

\maketitle


\newpage

\section{Introduction}

Quantum quench at finite quench rates which involve critical points are known to display universal scaling behavior in various regimes. For quench rates which are slow compared to physical mass scales local quantities in many systems obey Kibble Zurek scaling \cite{kibble,zurek,more,qcritkz}. In systems which have relativistic continuum limits there is a different scaling behavior for quench rates which are fast compared to physical mass scales, but slow compared to the UV scale \cite{dgm1,dgm2,dgm3,dgm4,ddgms,smolkin}. Finally at quench rates at the scale of a UV cutoff one expects that the response saturates as a function of the rate. These scaling behaviors are characteristic of early time response, i.e. for measurements made in the middle of the quench or soon after the quench is over. 

While the scalings themselves are quite generic, explicit investigations typically involve solvable models and a lot has been learnt from exactly solvable quench protocols in these models. They have also been studied for models which have holographic descriptions via the AdS/CFT correspondence \cite{Das:2016eao,holokz,fastholo}. In fact fast quench scaling was first discovered in holographic studies in \cite{fastholo}. They have been most extensively studied for local quantities like one point functions and correlation functions. For one dimensional harmonic chain scaling has also been found for the entanglement entropy \cite{Caputa:2017ixa} and recently for circuit complexity \cite{dd}.

Complexity in a field theory quantifies the difficulty in preparing a quantum state starting from some reference state. Study of such measures is motivated by ideas of holographic complexity \cite{susskind1}-\cite{hmjqo}. Since this is a quantity which characterizes properties of a quantum state which are not easily captured by correlation functions, it is interesting to study its behavior in non-equilibrium situations. There are several proposals for quantifying complexity in field theories. The proposal we consider in this paper is "circuit complexity" which relates the length of the optimal circuit of unitary operations relating the reference state and a target state to a geometric quantity in the space of states parametrized in a suitable fashion \cite{myers1}-\cite{yk}. Clearly, because of the dependence on the reference state as well as the unitary gates used, this is not uniquely defined. Nevertheless such a definition is expected to capture the true complexity of a state and seems to agree with holographic expectations.
For other approaches to field theoretic complexity, see \cite{othercomplex1, othercomplex2, othercomplex3, othercomplex5, othercomplex4, othercomplex6}.

In this paper we study scaling of circuit complexity in quantum quench for $1+1$ dimensional majorana fermions on a spatial lattice with a time (and momentum) dependent mass function - this is the fermionic description of a one dimensional transverse field Ising model with a time dependent transverse field. Following \cite{ddgms} we consider a time dependence for which the dynamics can be solved exactly - this corresponds to a transverse field which asymptotes to constant values at early and late times and passes through the critical point at some intermediate time which we choose to be $t=0$. The Heisenberg picture state of the system is chosen to be the "in" state, which approaches the ground state of the system at early times. This latter state is also chosen as the reference state. The Heisenberg picture state is then a Bogoliubov transformation of the reference state with time dependent Bogoliubov coefficients. 
As shown in \cite{myersfermion} for such a free fermion theory, circuit complexity (as defined in that paper) can be expressed entirely in terms of these Bogoliubov coefficients. Using the exact expression for this quantity we study the complexity analytically in various regimes. 

In the slow regime we use the standard adiabatic-diabatic sceneario underlying Kibble Zurek scaling to evaluate the complexity in the middle of the quench. We find a scaling behavior $\sim {\rm{constant}}+ (\dt)^{-1/2}$ where $\dt$ denotes the time scale of the quench.
We compare this result with a numerical evaluation of the integral involved in the exact result and find excellent agreement.
Interestingly this  comes mostly from contribution of modes which remain adiabatic. This is in contrast to what happens for the bosonic theory studied in \cite{dd} where the zero momentum modes in fact dominate the result. 

In the fast regime, we can perform an expansion of the exact answer in a power series in $Jb \dt$ where $J$ denotes the mass scale of the theory and $b$ is the quench amplitude. In this expansion, the complexity at $t=0$ is proportional to $\dt$ for arbitrarily small $\dt$. This agrees nicely with a numerical evaluation of the exact answer. The complexity at a slightly later time $t \ll \dt$ shows a slightly different behavior : for $\dt$ smaller than a $t$-dependent threshold value the complexity saturates as a function of $\dt$, while for $\dt$ larger than this threshold, the above mentioned linear behavior holds.

The content of the paper is as follows: In section \ref{cfft}, we summarize the definition of circuit complexity. In section \ref{mqd}, we introduce 1D Majorana fermion field theory and the derivation of complexity of its quench by considering Bogoliubov transformation. In section \ref{sc}, we study the scaling of complexity with respect to quench rate. In section \ref{dc}, we discuss the similarity and defference between 1D Majorana fermionic field theory and bosonic field theory in \cite{dd}, then we show some numerical results of the late-time behaviors of complexity. 

\section{Complexity in Free fermionic theory}
\label{cfft}

We follow the definiton of complexity in \cite{myers1} and \cite{myersfermion} which we summarize below: complexity is the minimal number of elementary unitary gates needed to prepare a certain target state $| \psi_T \rangle$ from a reference state $| \psi_R \rangle$
\begin{equation}
|\psi_T \rangle = U | \psi_R \rangle, U=\prod_{i=1}^{N} V_i .
\end{equation}
In continuum limit, $U$ takes a form of path-ordered exponential of the sum of products of control function $Y^I(s)$ and a basis of elementary gates $\mathcal{O}_I$
\begin{equation}
U(s) = \overleftarrow{\mathcal{P}} \text{exp}\left[ -i \int_{0}^s \text{d}s \sum_I Y^I(s) \mathcal{O}_I\right], U=U(s=1)
\end{equation}
And the complexity is defined to be the circuit that minimizes a cost 
\begin{equation}
\mathcal{D}(U(t)) = \int_0^1 \text{d}s F\left( U(s); Y^I(s) \right).
\end{equation}

Notice that $Y^I(s)$ can be intepreted as the $I^{th}$ component of the tangent vector of trajectory $U(s)$. The functional $F$ is a measurement of ``distance'' from reference state at $U(0)$ to target state at $U(1)$: for example, if all classes of gates have equal cost, $F$ can have a general form $F_{\kappa}(U;Y^I)=\sum_I |Y^I|^{\kappa}$. Then minimizing the cost is equivalent to looking for the shortest geodesic on the manifold formed by tangent vector $\vec{Y}(s)$. 

When both the target state $| \psi_T \rangle$ and the reference state $| \psi_R \rangle$ are gaussian, there exist two pairs of sets of creation and annihiliation operators $\{a_T\}, \{a_T^{\dagger}\}$ and $\{a_R\},\{ a_R^{\dagger}\}$ s.t. $a_{T} | \psi_{T} \rangle =0$ and $a_{R} | \psi_{R} \rangle =0$. Then the transformation between the two states can be described by the transformation between the two pairs of creation and annihiliation operators. Most of time the transformation is a Bogoliubov transformation
\footnote{ A single-fermion excited state $| k \rangle = a_k^{\dagger} | \text{vac}\rangle$ with fermion momentum $k$ can be expressed in the form ``$f_p | k \rangle =0, \forall p $'' as well, i.e. one can find sets of creation and annihilation operators, $\{ f_p \}$ and $\{ f_p^{\dagger} \}$, to represent $| k \rangle$ as their vacuum state. This is because vacuum state $| \text{vac} \rangle$ satisfies ``$a_p | \text{vac} \rangle =0, \forall p$'', and as a result one can always define $f_k \equiv a_k^{\dagger}$ (while for other $p$, $f_p \equiv a_p$). This implies that single-fermion excited state is gaussian. However, if we choose single-fermion excited state as the target state and vacuum state as the reference state, there is no Bogoliubov transformation between these two gaussian states.
}. Below we give a rudimentary argument about fermions:

For a pair of fermions, the unitary operation $U$ from reference state $| \psi_R \rangle$ to target state $| \psi_T \rangle$ is of the form
\begin{equation}
\begin{split}
\tilde{a} = \alpha a - \beta b^{\dagger}, \hfill \\
\tilde{b}^{\dagger} = \alpha^* b^{\dagger} + \beta^* a, 
\label{bogo_t}
\end{split}
\end{equation}
where operators $a, b$ and $a^{\dagger}, b^{\dagger}$ are annihilation and creation operators of reference state, i.e. $ a| \psi_R \rangle = b| \psi_R \rangle = 0$; Similarly, $\tilde{a}, \tilde{b}$ and $\tilde{a}^{\dagger}, \tilde{b}^{\dagger}$ are annihilation and creation operators of target state, i.e. $ \tilde{a}| \psi_T \rangle = \tilde{b}| \psi_T \rangle = 0$. To preserve the anti-commutation relations, $\alpha$ and $\beta$ satisfy
\begin{equation}
|\alpha|^2 + |\beta|^2=1.
\label{al_be}
\end{equation}

The equation (\ref{al_be}) implies that all of the possible target states form a unit sphere with the north pole the reference state. This is made explicit by writing $\alpha$ and $\beta$ by two angles $\theta$ and $\phi$, i.e.
\begin{equation}
\alpha=\text{cos} \theta, \beta=e^{i\phi}\text{sin} \theta.
\label{theta}
\end{equation} 
Then a definition of the circuit complexity is the length of the geodesic from north pole to the position of the target state, i.e. $|\theta|$ gives the minimal cost. This can be generalized to N-pairs of free fermions. Since the Bogoliubov transformation does not mix operators with different momenta, it still takes the form in (\ref{bogo_t}) and therefore (\ref{theta}) for each pair of fermion with momentum $\vec{k}, -\vec{k}$. On the other hand, to prepare the target state $| \psi_T \rangle $ from the reference state $| \psi_R \rangle $, one need to Bogoliubov transform all the independent (momentum) modes. As a result, the circuit complexity is the sum of geodesics $|\theta|(\vk)$ of all momenta, i.e.

\begin{equation}
\mathcal{C}^{(n)} =\sum_{\vk} |\theta|^n(\vk) \to {V \int \frac{\text{d}^d k}{(2\pi)^d} |\theta|^n(\vec{k})},
\label{1-1}
\end{equation}
where
\begin{equation}
\left| \theta \right|(\vec{k}) = \text{tan}^{-1} \frac{|\beta_{\vec{k}}|}{|\alpha_{\vec{k}}|} 
= \frac{1}{2}\text{tan}^{-1} \frac{2|\alpha_{\vec{k}}||\beta_{\vec{k}}|}{\left||\alpha_{\vec{k}}|^2-|\beta_{\vec{k}}|^2\right|}.
\label{cmpl_def}
\end{equation}

\section{The model and quench dynamics}
\label{mqd}

The model considered in this paper is Majorana fermion field theory of the one dimensional transverse field Ising model with a time dependent transverse field (The model is discussed in details in \cite{ddgms}). The Hamiltonian is given by
\begin{equation}
H=\int \frac{{\text d} k}{2\pi} \chi^{\dagger}(k,t) \left[ -m(k,t)\sigma_3 + G(k)\sigma_1 \right] \chi(k,t).
\label{tdH}
\end{equation}
where $\sigma_{1,3}$ are 2D Gamma matries and $\chi$ denotes the two component spinor field, i.e. $\chi=\left( {\begin{array}{*{20}{c}}
 \chi_1(k) \\
 \chi_2(k) \\
\end{array} } \right)$.

The Heisenberg equation of motion for $\chi(k,t)$ is a superposition of two independent solutions  $U(k,t)$ and $V(k,t)$,
\begin{equation}
i \partial_t \left(U(k,t),V(-k,t)\right)= \left[ -m(k,t)\sigma_3 + G(k) \sigma_1 \right] \left(U(k,t),V(-k,t)\right)
\label{heis_eom}
\end{equation}
and
\begin{equation}
\chi({k},t)=a({k})U({k},t)+a^{\dagger}(-{k})V(-{k},t).
\end{equation}
because of Majorana condition $\chi_2(k)=\chi_1^{\dagger}(-k)$.
The operators $a({k})$ and $a^{\dagger}({k})$ satisfy the usual anti-commutation relations
\bea
\{a (k) , a^\dagger (k') \}  =& \delta (k-k') & \nonumber \\
\{a(k), a(k')\}  =  & \{ a^{\dagger}(k), a^{\dagger}(k') \} & = 0
\eea

We can relate the spinor to a scalar field $\phi(k,t)$ by letting
\begin{equation}
\begin{split}
U({k},t)= \left( {\begin{array}{*{20}{c}}
 -i\partial_t +m({k},t) \\
 -G({k}) \\
\end{array} } \right) \phi({k},t), \hfill \\
V(-{k},t) = \left( {\begin{array}{*{20}{c}}
 G({k}) \\
  i\partial_t +m({k},t) \\
\end{array} } \right)\phi^*({k},t),
\label{1-2}
\end{split}
\end{equation}
where $\phi(k,t)$ satisfies 
\begin{equation}
\partial_t^2 \phi +i\partial_t m \cdot \phi +(m^2+G^2)\phi=0.
\label{heis_eom2}
\end{equation}
according to (\ref{heis_eom}), and 
\begin{equation}
\left| \partial_t \phi \right|^2+(m^2+G^2)\left|\phi \right|^2-2m\cdot \text{Im} \left(\phi \partial_t \phi^* \right)=1 
\label{acommu2}
\end{equation}
to preserve anti-commutation relations and the orthonormality of $U(k,t)$ and $V(k,t)$.

An exactly solvable quench dynamics has been found in \cite{ddgms} which we use
\begin{equation}
m(k,t)=A(k)+B\text{tanh}(t/\delta t),
\label{mass}
\end{equation}
and the rest of the parameters are 
\begin{equation}
A(k)=2J(a-\text{cos}k), B=2Jb, G(k)=2J\text{sin}k,
\end{equation}
where $J$ is the interaction strength between the nearest-neighbor spins in Ising model. It has dimension of energy. $a$ is the lattice spacing of Ising model. $b$ determines the mass gap.

In the rest of the paper we mainly consider the case $a=1$, which describes a cross-critical-point (CCP) type-like potential at $k=0$; another interesting case is when $a=1-b$, which corresponds an end-critical-point (ECP) type-like potential at $k=0$. 

The Heisenberg picture state we use is the "in" state. This means that
the spinors $U(k,t)$ should asymptote to the positive frequency solution of the equation in the infinite past. This "in" solution is given by (\ref{heis_eom2}) and (\ref{acommu2}) is 
\begin{equation}
\begin{split}
\phi_{in}({k},t ) =& \frac{1}{|G(k)|}\sqrt{\frac{\omega_{in}+m_{in}}{2\omega_{in}}}\text{exp}[-i\omega_+({k})t-i\omega_-({k})\delta t \text{log}(2 \text{cosh}(t/\delta t))] \hfill \\
& _2F_1[1+i\omega_-({k})\delta t+iB \delta t, i\omega_-({k})\delta t - i B \delta t; 1-i\omega_{in}({k})\delta t; \frac{1}{2}(1+\text{tanh}(t/\delta t ))],  \hfill \\
\end{split}
\label{f_sol}
\end{equation}
where the frequencies $\omega_{in, out, \pm}$ are defined to be
\begin{equation}
\begin{split}
\omega_{out,in}=\sqrt{G({k})^2+(A({k})\pm B)^2}, 
\omega_{\pm}=\frac{1}{2}(\omega_{out}\pm \omega_{in}). \hfill \\
\end{split}
\end{equation}
and $m_{in}= m(t \to -\infty) = A(k)-B$.

For the reference state we will choose the ground state of the system in infinite past, while the target state is the Heisenberg picture state. For some momentum $k$ the former is annihilated by a set of fermionic oscillators $a_{-\infty}({k})$ and $a^{\dagger}_{-\infty}({k})$ defined by
\ben
\chi(k,t)=a_{-\infty}(k)U_{-\infty}({k},t)+a^{\dagger}_{-\infty}(-k)V_{-\infty}(-{k},t)
\een
where $U_{-\infty}(k,t)$ and $V_{-\infty}(k,t)$ are given by the expressions (\ref{1-2}) using the asymptotic form $\phi_{-\infty}(k,t)$
\ben
\phi_{-\infty}(k,t) = \frac{1}{\sqrt{2\omega_{in}(\omega_{in}-m_{in})}}e^{-i\omega_{in}t}
\een
The relationship between $a_{-\infty}({k}),a^{\dagger}_{-\infty}({k})$ and 
$a(k),a^{\dagger}(k)$ then becomes a set of Bogoliubov transformations of the form
(\ref{bogo_t}) 
\begin{equation}
\begin{split}
a_{-\infty}({k},t)=\alpha({k},t)a({k})-\beta({k},t)a^{\dagger}(-{k}), \hfill \\
a_{-\infty}^{\dagger}(-{k},t)=\beta^*({k},t)a({k})+\alpha^*({k},t)a^{\dagger}(-{k}), \hfill \\
\end{split}
\label{bogo_t2}
\end{equation}
where $\alpha(k,t)=\alpha(-k,t)$ and $\beta(k,t)=-\beta(-k,t)$; anti-commutation relation requires $|\alpha(k,t)|^2+|\beta(k,t)|^2=1$.
Then
$\alpha(k,t)$ and $\beta(k,t)$ can be expressed by $U(k,t), V(k,t)$:
\begin{equation}
\begin{split}
\alpha({k},t)=U_{-\infty}^{\dagger}({k},t)U({k},t)=V^{\dagger}(-{k},t)V_{-\infty}(-{k},t), \hfill \\
\beta({k},t)=-U_{-\infty}^{\dagger}({k},t)V(-{k},t)=U^{\dagger}({k},t)V_{-\infty}(-{k},t), \hfill \\
\end{split}
\label{albe_2}
\end{equation}
and therefore $\phi(k,t)$:
\begin{equation}
\begin{split}
\alpha({k},t)=\phi_{-\infty}^*({k},t)\left\{ G^2({k})+(-\omega_{in}+m_{in})\left[ -i\partial_t +m({k},t)\right] \right\} \phi({k},t) \hfill \\
\beta({k},t)=\phi_{-\infty}^*({k},t)G({k})\left\{\left[ i\partial_t +m({k},t)\right] - (-\omega_{in}+m_{in})\right\} \phi^*({k},t). \hfill \\
\end{split}
\label{albe}
\end{equation}

In this paper we consider the measurement closest to the original definition of complexity in the discrete case, i.e. $F(U,\vec{Y})=\sum_I \left| Y^I \right|$, therefore the circuit complexity of the model is
\begin{equation}
\mathcal{C}^{(1)} ={V \int \frac{\text{d} k}{2\pi} |\theta|({k},t)},
\end{equation}
where the integrand $\theta(k,t)$ is given by
\begin{equation}
\left| \theta \right|({k},t) \equiv \text{tan}^{-1} \frac{|\beta({k},t)|}{|\alpha({k},t)|} = \tan^{-1} \left| \frac{\phi_{-\infty}({k},t)G({k})\left\{\left[ -i\partial_t +m({k},t)\right] - (-\omega_{in}+m_{in})\right\} \phi({k},t)}{\phi_{-\infty}^*({k},t)\left\{ G^2({k})+(-\omega_{in}+m_{in})\left[ -i\partial_t +m({k},t)\right] \right\} \phi({k},t)} \right|
\end{equation}
For practical reason, we ignore factor $\frac{V}{2\pi}$. 


\section{Scaling of Complexity}
\label{sc}

\begin{figure}
\centering
\includegraphics[width=0.7\textwidth]{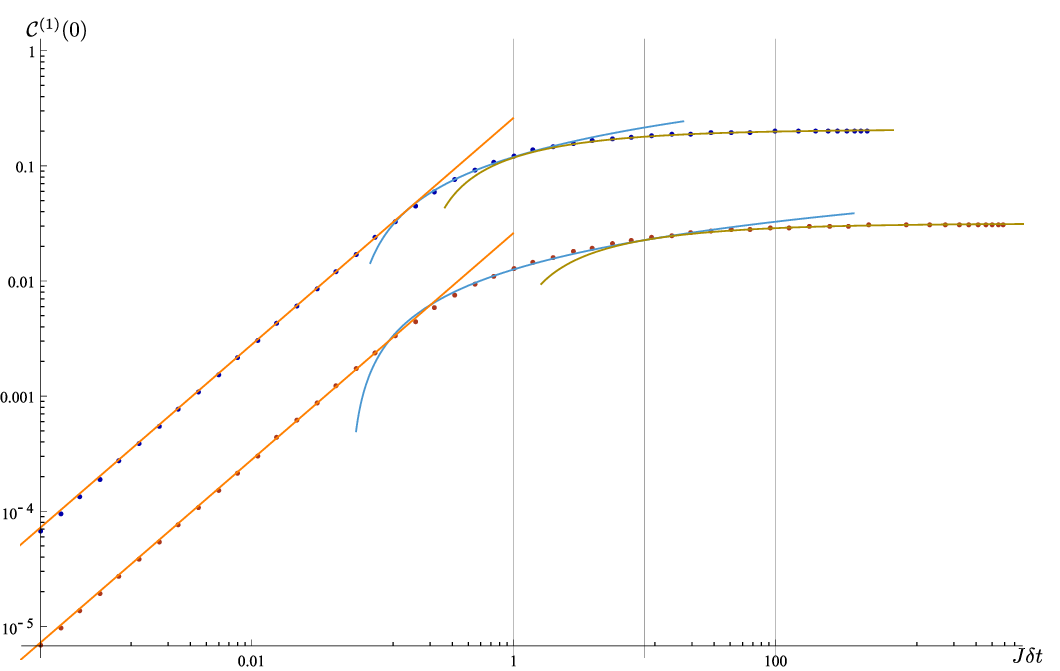}
\caption{Exact $\mathcal{C}^{(1)}(0)$-$\delta t$ relations in log-log scale. Red and blue dots correpond to $b=0.01$ and $b=0.1$ respectively. The orange, blue, and yellow fitting curve are $y=cx^d$, $y=P+Q\log x$, and $y=P'+Q'x^{-1/2}$, respectively. The linear fitting coefficient $d=0.985146$ for $b=0.1$ and $d=0.984975$ for $b=0.01$, which implies the linear relation between $\mathcal{C}^{(1)}(0)$ and $\delta t$ in fast quench regime.}
\label{f9}
\end{figure}

\begin{figure}
\centering
\includegraphics[width=0.7\textwidth]{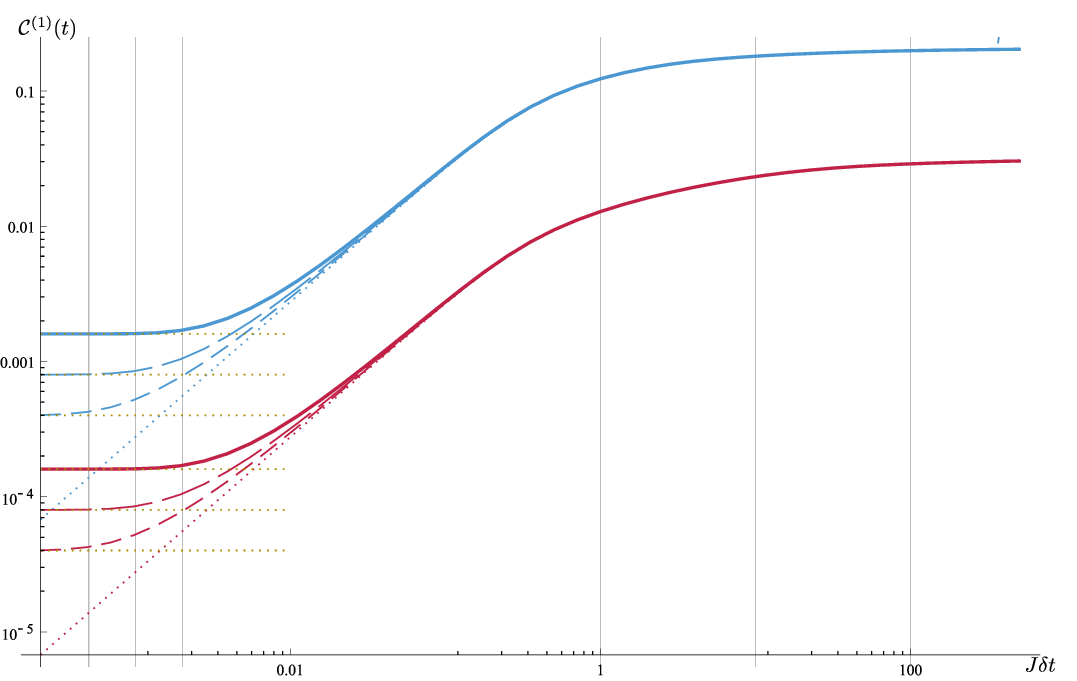}
\caption{Exact $\mathcal{C}^{(1)}(t)$-$\delta t$ relations in log-log scale. Red and blue lines correpond to $b=0.01$ and $b=0.1$ respectively. From solid to dashed, the curves correspond to $t=0.002, 0.001$ and $0.0005$, respectively. We can see the circuit complexity saturates around $\delta t \sim t$ (gridlines), and the saturation value is approximately $8Jbt$ (in yellow dotted lines). As reference, $\mathcal{C}^{(1)}(0)$-$\delta t$ relations are in dotted lines.}
\label{f10}
\end{figure}

Now we want to see how circuit complexity scales with respect to quench rate.  The behavior of  $\mathcal{C}^{(1)}(t)$ as a function of  $\delta t$ is shown in Fig. \ref{f9} and Fig. \ref{f10}. There are three regimes of the quench rate: slow quench, fast quench and instantaneous quench.


\begin{figure}
\centering
\subfigure[]{ \label{f6}
\includegraphics[width=0.45\textwidth]{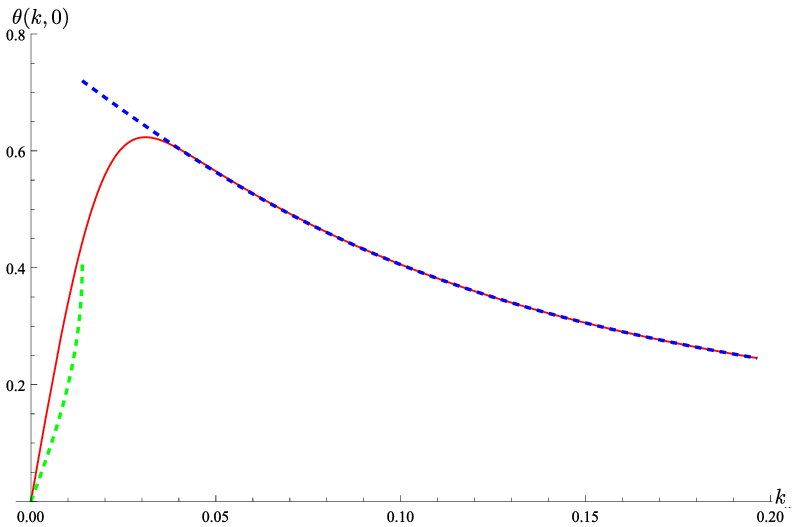}}
\subfigure[]{ \label{f6b}
\includegraphics[width=0.45\textwidth]{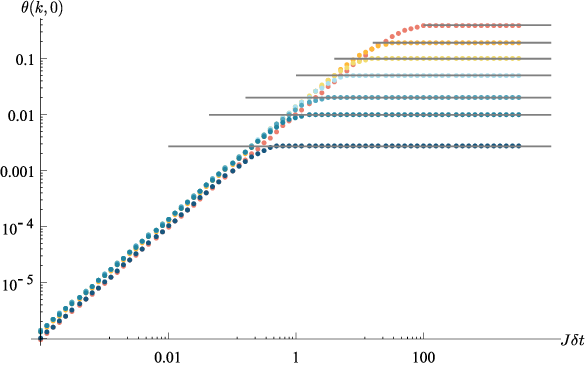}}
\caption{\ref{f6} Relation between single-mode contribution $|\theta|(k,0)$ and momentum $k$: Red line shows the exact mode contribution to complexity $\theta(k,0)$; green and blue dashed lines are approximated complexity with KZ mass and critical mass, respectively. $b=0.1$. \ref{f6b} Relation between $|\theta|(k,0)$ and $J\delta t$: From dark blue to red, $k=1.5,0.5,0.25,0.1,0.05,0.025,0.01$; the grey solid horizontal lines are adiabatic approximations (\ref{tt_sq}) when $J\delta t > b\csc^2 k$. $b=0.01$.}
\label{f6ab}
\end{figure}

\subsection{Slow quench} \label{ssq}
In slow quench region, $J \delta t \gg 1/b$, the asymptotic behavior of circuit complexity with respect to the quench rate is $\mathcal{C}^{(1)} \sim P'+ Q'\delta t^{-1/2}$, where $P',Q'$ are constants (Fig. \ref{f9}). 
This behavior is consistent with Kibble-Zurek scaling. 
In particular, the system evolves adiabatically at the beginning because the rate of change of time-dependent coupling ($m(k,t)$ in this system) is much smaller than the square of the initial energy gap. However, the adiabaticity breaks down as one approaches the critical point at a time called Kibble-Zurek time $t_{KZ}$, because of critical slowing down. Since the instantaneous energy gap scales with time-dependent coupling $m(k,t)$ with the correlation length exponent $\nu$ as $E_g \sim |m(k,t)-m_c|^{\nu}$. $t_{KZ}$ can be found by using the Landau criterion
\begin{equation}
\frac{1}{E_g^2(t)} \frac{\text{d}E_g(t)}{\text{d}t} \bigg|_{-t_{KZ}} \sim 1,
\label{lc}
\end{equation}
where $E_g$ is the instantaneous energy gap, and in the model we study,
\begin{equation}
E_g = \sqrt{m(k,t)^2+G(k)^2}.
\label{ener_gap}
\end{equation}
Adiabaticity holds when the quantity is much smaller than 1. 

Clearly the Kibble-Zurek time depends on the momentum of the mode according to (\ref{lc}) and (\ref{ener_gap}). In CCP type-like potential case, one can find that adiabaticity breaks down at $-t_{KZ}$ and reappears at $t_{KZ}$. On the other hand, in ECP type-like potential case, adiabaticity breaks down at $t_{KZ}$.

To make it clear, the circuit complexity is plotted mode by mode at $t=0$ (See Fig. \ref{f6}). We find that the momentum-dependence of circuit complexity can be divded into two regions by Landau criterion. 
In particular, a critical momentum $k_c$ exists, where 
\begin{equation}
\text{sin} k_c = \sqrt[4]{\frac{1}{27}}\sqrt{\frac{b}{J\delta t}}.
\label{k_cond}
\end{equation}
For $k < k_c$ the adiabatic approximation breaks down, and the system is frozen after $-t_{KZ}$. This implies that circuit complexity can be evaluated approximately by that at a fixed mass $m_{KZ}=m(-t_{\text{KZ}})$, which can be treated as an effective mass of the frozen system. For $k > k_c$, adiabatic approximation is valid, and circuit complexity can be evaluated by using the instantaneous mass at $t=0$. Then in general, the approximation is
\begin{equation}
|\theta| \approx \frac{1}{2} \text{tan}^{-1} \left| \frac{(m_{in}-m)G}{m_{in}\cdot m+G^2} \right|,
~~m= \left\{ {\begin{array}{*{20}{c}}
 m_{KZ}, & k< k_c\\
 m(t=0), & k>k_c \\
\end{array} } \right. .
\label{tt_sq}
\end{equation}
Therefore, by summing over all momenta ($k \in [0,\pi]$), an approximation of circuit complexity is 
\begin{equation}
\begin{split}
\mathcal{C}^{(1)}(0)\approx & \frac{1}{2} \int_0^{\pi}\text{d}k \text{tan}^{-1} \left| \frac{[m_{in}-m(t=0)]G}{m_{in}\cdot m(t=0)+G^2} \right| \hfill \\
&-\frac{1}{2} \int_0^{k_c}\text{d}k \text{tan}^{-1} \left| \frac{[m_{in}-m(t=0)]G}{m_{in}\cdot m(t=0)+G^2} \right| +\frac{1}{2} \int_0^{k_c}\text{d}k \text{tan}^{-1} \left| \frac{[m_{in}-m_{\text{KZ}}]G}{m_{in}\cdot m_{\text{KZ}}+G^2} \right|,
\end{split}
\end{equation}
where the 1st and the 2nd terms are the adiabatic approximation of $|\theta|$ for $k> k_c$, and the 3rd term is the response of the system at a constant mass equal to the Kibble-Zurek mass $m_{KZ}$ when adiabaticity breaks down. 
Under the condition $J \delta t \gg 1/b$, we can simplify the complexity and determine the leading and subleading term of it:
\begin{equation}
\mathcal{C}^{(1)}(0) \sim \int_{0}^{\pi/2}\text{d}x \text{tan}^{-1} \left( \frac{b}{2-b}\text{cot} x \right) + \mathcal{O}\left(\sqrt{\frac{b}{J\delta t}}\right).
\label{cc_sq}
\end{equation}
Numerically we can find that leading term is 0.207762 when $b=0.1$, which is close to the curve fitting result of exact $\mathcal{C}^{(1)}(0)$, $P'=0.207605$ (yellow fitting in Fig. \ref{f9}). Therefore, in the slow quench regime, circuit complexity follows the expectation of KZ.

\subsection{Instantaneous quench}
In the instantaneous quench regime, i.e. $J\delta t \ll 1$, we find that circuit complexity at time $t$ shows linearity when $\delta t \gg t$ and saturation when $\delta t \ll t$; In both cases the circuit complexity is proportional to the gap of mass $Jb$ (Fig. \ref{f10}). One can see these features more clearly from the asymptotic behaviors of $\phi$ and $\beta$ when $t, \delta t \ll 1$. 

Notice that when $\delta t \gg t$, $\text{tanh}(t/\delta t) \to t/\delta t$. Therefore we can expand solution $\phi_{in}(k,t)$ in (\ref{f_sol}) and find
\begin{equation}
\begin{split}
\phi_{in}(\vec{k},t ) 
\approx & \frac{1}{|G(k)|}\sqrt{\frac{\omega_{in}+m_{in}}{2\omega_{in}}}e^{-i\omega_+t} \hfill \\
&\times \left\{ 2^{-i\omega_-\delta t} \frac{\omega_{+}+B}{\omega_{out}}[1- i\delta t(\omega_--B) \text{log}(\frac{1}{2}(1+\frac{t}{\delta t}))]  \right. \hfill \\
&\left.+2^{i\omega_+\delta t}(1-\frac{t}{\delta t})^{-i\omega_{out}\delta t}\frac{\omega_--B}{\omega_{out}}[1+i\delta t(\omega_++B) \text{log}(\frac{1}{2}(1+\frac{t}{\delta t}))] \right\}  \hfill \\
& -\frac{1}{|G(k)|}\sqrt{\frac{\omega_{in}+m_{in}}{2\omega_{in}}}e^{-i\omega_+t} \times t \delta t{(\omega_{+}+B)(\omega_--B)} \text{log}2 \hfill \\
& + (\text{higher order contributions}).
\end{split}
\end{equation}
Then the leading term of $\beta^*(k,t)$ (\ref{albe}) is now approximately
\begin{equation}
\begin{split}
\beta^*(k,t) \approx  &  \frac{1}{G}{\frac{\omega_{in}+m_{in}}{2\omega_{in}}}e^{-i\omega_+t-i\omega_{in}t}
 \left\{i2\delta t (\omega_{+}+B)(\omega_--B)\text{log}2  \right\}.  \hfill \\
\end{split}
\label{tt_iq}
\end{equation}
Since $J\delta t \ll 1$, circuit complexity $|\theta(k,0)| \approx |\beta|$ and as a result, we can plug $G(k)$ in and estimate the complexity to be
\begin{equation}
\mathcal{C}^{(1)}(0) \sim bJ\delta t  \cdot 4\text{log}2 \approx 2.77bJ\delta t. 
\end{equation}
when $b \ll 1$.
This is close to the linear fitting of exact circuit complexity in Fig. \ref{f9}, where the slopes are $c \approx 0.261913$ when $b=0.1$ and $c \approx 0.0261736$ when $b=0.01$.

When $\delta t \ll t$, $\text{tanh}(t/\delta t) \to 1-e^{-2t/\delta t}$. Again one can expand solutions and find
\begin{equation}
\begin{split}
\phi_{in}(\vec{k},t ) 
\approx & \frac{1}{|G(k)|}\sqrt{\frac{\omega_{in}+m_{in}}{2\omega_{in}}}\text{exp}[-i\omega_-\delta t e^{-2t/\delta t}] \hfill \\
&\times \left\{e^{-i\omega_{out}t}\frac{\omega_{+}+B}{\omega_{out}}\left( 1+i\delta t(\omega_--B)e^{-2t/\delta t} \right) 
+e^{i\omega_{out}t}\frac{\omega_--B}{\omega_{out}}\left(1-i\delta t(\omega_++B)e^{-2t/\delta t} \right) \right\}.  \hfill \\
\end{split}
\end{equation}
Therefore $\beta(k,t)^*$ is approximately
\begin{equation}
\begin{split}
\beta^*(k,t) \approx &-i\text{exp}[-i\omega_-\delta t e^{-2t/\delta t}]\frac{G(m_{out}-m_{in})}{\omega_{in}\omega_{out}}e^{-i\omega_{in}t}\text{sin}\omega_{out}t \hfill \\
&+\frac{1}{G}{\frac{m_{in}+\omega_{in}}{2\omega_{in}}}e^{-i\omega_{in}t}\text{exp}[-i\omega_-\delta t e^{-2t/\delta t}]4(\omega_--B)e^{-2t/\delta t}e^{i\omega_{out}t},\hfill \\
\end{split}
\end{equation}
and then the leading term and subleading term of the circuit complexity is
\begin{equation}
\mathcal{C}^{(1)}(t) \sim 8bJt + \mathcal{O}(te^{-2t/\delta t}).
\end{equation}
Thus the leading term is $\delta t$-independent, and linearly increase as $t$ increases, which corresponds to the saturation when $\delta t \ll t$ in Fig. \ref{f10}.

Behavior of circuit complexity in instantaneous quench regime is consistent with the behavior when quench occurs instantaneously ($\delta t  \to 0$). In the latter case the time-dependent mass in (\ref{mass}) can be described by
\begin{equation}
m(k,t)= A(k) + B \left( \Theta(t/\delta t) -1/2 \right)
\end{equation}
where $\Theta(x)$ is the Heaviside step function. The circuit complexity can be exactly figured out. The exact ``in'' solution is
\begin{equation}
\phi(t)=\frac{1}{|G|}\sqrt{\frac{m_{in}+\omega_{in}}{2\omega_{in}}} \times \left\{ {\begin{array}{*{20}{c}}
e^{-i\omega_{in}t} , & t<0\\
\frac{\omega_++B}{\omega_{out}}e^{-i\omega_{out}t}+\frac{\omega_--B}{\omega_{out}}e^{i\omega_{out}t}. & t \ge 0\\
\end{array} } \right.  \hfill \\
\label{e16}
\end{equation}
and thus the circuit complexity 
\begin{equation}
\mathcal{C}^{(1)}(t) = \int_0^{\pi} \text{d} k  \frac{2b \text{sin}k}{\sqrt{\text{sin}^2k+(1-\text{cos}k-b)^2}}\cdot 2Jt \approx \frac{8bJt}{\sqrt{1-b}} \approx 8bJt
\end{equation}
when $b \ll 1$.

One can see that at $t=0$ the circuit complexity $\mathcal{C}^{(1)}(0)$ does not saturate. This is different from $\langle \bar{\chi} \chi \rangle$, which saturates when $J \delta t < |b|$ (\cite{ddgms}).

\subsection{Fast quench}

Between slow quench $J \delta t \ll 1$ and instantaneous quench $J \delta t \gg 1/b$, we find a logarithmic dependence, i.e.
$\mathcal{C}^{(1)}(0) \sim P + Q\log(J\delta t)$, where $P, Q$ roughly linearly rely on $b$. 

One can easily generalize the other definition of complexity in (\ref{cmpl_def}) to the time-dependent case, s.t.
\begin{equation}
 \left| \theta \right|({k},t) \equiv \frac{1}{2}\text{tan}^{-1} \frac{2|\alpha({k},t)||\beta({k},t)|}{\big| |\alpha({k},t)|^2-|\beta({k},t)|^2 \big|}
 \label{4-3-1} 
 \end{equation}
Combined with ((\ref{1-2}) and (\ref{albe_2})), we can rewrite the denominator $\big| |\alpha({k},t)|^2-|\beta({k},t)|^2 \big|$ in terms of the c-number $\bar{V}V \equiv V(-k,t)^{\dagger}\sigma_3 V(-k,t)$:

\begin{equation}
|\alpha({k},t)|^2-|\beta({k},t)|^2 = \frac{m_{in}}{\omega_{in}} \bar{V}V + \frac{G}{\omega_{in}}\cos \gamma \sqrt{1-(\bar{V}V)^2}
\end{equation}
where $\gamma$ is the angular part of $-2G(k) \phi^*(-i \partial_t +m(k,t))\phi$.

On the other hand, the quench Hamiltonian (\ref{tdH}) can be rewriten into the form
\begin{equation}
\begin{split}
H =& \int \frac{{\text d} k}{2\pi} \chi^{\dagger}(k,t) \left[ -m_{in}\sigma_3 + G(k)\sigma_1 \right] \chi(k,t) - \int \frac{{\text d} k}{2\pi} \delta m(t) \bar{\chi}(k,t) \chi(k,t)  \hfill \\
\equiv & H_{CFT} - \delta \lambda \int {\text d} x  F(t/\delta t) \mathcal{O}_{\Delta},
\end{split}
\end{equation}
where $\delta \lambda = 2B$, $F(t/\delta t) = (1+\tanh(t/\delta t))/2$, and $\mathcal{O}_{\Delta} =  \bar{\chi}(x.t)\chi(x,t) $ is an operator with comformal dimension $\Delta$. Therefore one can use the Kubo formula to find the leading terms of $\langle \mathcal{O}_{\Delta}(t) \rangle \equiv \langle 0| \mathcal{O}_{\Delta}(t) |0 \rangle $ when $t \ll \delta t$, where state $|0 \rangle$ is the ``in'' vacuum that satisfies $a(k)|0\rangle =0$ for all momentum $k$:
\begin{equation}
\langle \mathcal{O}_{\Delta}(x,t) \rangle = \langle \mathcal{O}_{\Delta}(x,t) \rangle_{m_{in}} - \delta \lambda \int_{-\delta t}^t {\text d} t' F(t'/\delta t) \int \text{d} x' G_{R,m_{in}}(x,t;x',t') + ...
\end{equation}
where $G_{R,m_{in}}(x,t;x',t')$ is the retarded Green's function
\begin{equation}
G_{R,m_{in}}(x,t;x',t') = i \Theta(t) \langle [ \mathcal{O}_{\Delta}(x,t), \mathcal{O}_{\Delta }(x',t') ] \rangle_{m_{in}}
\end{equation}

The argument in \cite{Das:2016eao} \footnote{\cite{Das:2016eao} considered a general continuous field theory instead of a lattice one. Here we follow the logic and argue in continuous case as well.} implies that one can use the Green's function in CFT as an approximation to the exact one when $\delta t \ll (\delta \lambda)^{-\frac{1}{d-\Delta}}, (m_{in})^{-\frac{1}{d-\Delta}}$, where $d$ is the spacetime dimension. This is based on two facts: One is that the Green's function should satisfy causality. As a result the bound of integral over $x$ is actually $(-t+t', t-t')$. The other is that when space and time scales are both much smaller than the correlation length $(m_{in})^{-\frac{1}{d-\Delta}}$, the commutator in CFT is a good approximation to the Green's function. The argument leads to the conclusion that the integrals in the Kubo formula scale with $\delta t$ only, i.e.
\begin{equation}
\langle \mathcal{O}_{\Delta}(x,t) \rangle - \langle \mathcal{O}_{\Delta}(x,t) \rangle_{m_{in}}  \sim \delta \lambda (\delta t)^{d-2\Delta}
\end{equation}
A more concrete calculation in \cite{ddgms} shows 
\begin{equation}
\langle \bar{\chi}(x,t)\chi(x,t) \rangle - \langle \bar{\chi}(x.t)\chi(x,t) \rangle_{m_{in}} = \lim_{\eta \to 0} 2\pi \delta \lambda \log(\delta t/\eta)+...,~~~ \delta t \ll \frac{1}{\delta \lambda}, \frac{1}{m_{in}}
\end{equation}

Now, notice that $\bar{V}(-k.t)V(-k,t)$ is related to $\bar{\chi}_{n}(t)\chi_n(t)$ by a Fourier transform
\begin{equation}
\begin{split}
\langle 0 | \bar{\chi}_{n}\chi_n | 0 \rangle (t)  =\int_{-\pi}^{\pi} \frac{{\text d} k}{2\pi} e^{-in(k-k')} \langle 0 | \bar{\chi}(k,t)\chi(k',t) |0 \rangle = \int_{-\pi}^{\pi} \frac{{\text d} k}{2\pi} \langle 0 | \bar{V}(-k,t)V(-k,t) |0 \rangle, 
\end{split}
\end{equation}
on the Ising model we consider. Momentum $k \in (-\pi, \pi]$ is independent from the scale $\delta \lambda$ and $\delta t$. Thus the order of $\delta \lambda$ and $\delta t$ in the leading term and subleading term should not change. Figure out the leading term by plugging the $\phi_{-\infty}$ in, and we find 
\begin{equation}
\bar{V}V=\langle 0 | \bar{V}(-k,t)V(-k,t) |0 \rangle \approx \frac{m_{in}}{\omega_{in}} + \mathcal{O}(b \log(\delta t/\eta))
\end{equation}
and leading term of $\cos \gamma$ is 1.

Therefore, (\ref{4-3-1}) turns into
\begin{equation}
|\theta|(k,t) \approx \frac{1}{2} \tan^{-1} \frac{\mathcal{O}(b\log \delta t/\eta)}{1-\frac{1}{2}\left[ \mathcal{O}(b\log \delta t/\eta) \right]^2} 
\sim b\log (\delta  t/\eta)
\end{equation}
Circuit complexity is therefore of the form $P+Q \log (J\delta t)$ as Fig. \ref{f9} shows.

Here we need to make a comment on the higher order terms of $\cos \gamma$. In the fast quench regime, it can be ignored because $b \log(\delta t)$ is the second lowest order. However, in instantaneous regime, subleading term of $\cos \gamma$ is at the order $\mathcal{O}(\delta t)$, which can be found by expanding the Hypergeometric function. This might explain why $\mathcal{C}^{(1)}(t)$-$\delta t$ relation is not quadratic, though
\begin{equation}
\bar{V}V \approx \frac{m_{in}}{\omega_{in}} + \mathcal{O}((J\delta t)^2)
\end{equation}
when $J \delta t \ll 1$ (\cite{ddgms}).

\section{Discussions} \label{dc}

In many ways the circuit complexity $\mathcal{C}^{(1)}(0)$ of relativistic fermionic Ising theory scales in a way similar to free bosonic oscillators \cite{dd}. In particular, like circuit complexity of free bosonic oscillators when $\omega_0 \delta t \ll 1$, circuit complexity of free fermionic oscillators also shows linear behavior at $t=0$  in instantaneous quench regime ($J \delta t \ll 1$). This is because the contribution from each single (momentum) mode, $|\theta|(k,0)$, scales linearly as $\delta t$ varies (\ref{tt_iq}). To explain the behavior of circuit complexity $\mathcal{C}(0)$ in bosonic theory when slow quenched, \cite{dd} analyzed the contributions from single modes $C^k$ and found that $C^k$ scales logarithmically when $\delta t < \frac{\omega_0}{4} \csc^2 (\frac{k}{2})$ and saturates when $\delta t \ge \frac{\omega_0}{4} \csc^2 (\frac{k}{2})$. Here we can draw a similar conclusion from (\ref{k_cond}) that single nonzero mode contribution $|\theta|(k \ne 0,0)$ saturates when $J\delta t \gg b\csc^2 k$ (See Fig. \ref{f6b}). 
To be concrete, $k \gg k_c$ is the condition for $|\theta|(k,0)$ to be adiabatic. Then by plugging in (\ref{k_cond}) one can find the inequality between $k$ and $\delta t$
\begin{equation}
\sin k \gg \sin k_c =\sqrt[4]{\frac{1}{27}}\sqrt{\frac{b}{J\delta t}} \Longrightarrow J \delta t \gg \sqrt{\frac{1}{27}}b \csc^2 k \sim b\csc^2 k.
\end{equation}
Given that adiabatic result (\ref{tt_sq}) is independent of $J \delta t$ due to $m(t=0) = A(k)$, when the inequality is satisfied, $|\theta|(k,0)$ is independent of $\delta t$.
\footnote{ Here we cannot give a stronger condition such as $J \delta t \ge \sqrt{\frac{1}{27}}b \csc^2 k$ because according to Fig. \ref{f6}, adiabatic result does not match exact one when $J \delta t$ is slightly larger than $\sqrt{\frac{1}{27}}b \csc^2 k$.}

However, the zero mode contribution $C^{k=0}$ in free bosonic theory is very different from $|\theta|(k=0,0)$ in free fermionic theory.
In the free bosonic case, the contribution from zero-mode $C^{k=0} \sim \log \delta t$ in KZ regime due to the fact that the saturation happens when $\delta t \ge \frac{\omega_0}{4} \csc^2 (\frac{k}{2}) \to \infty$. 
In the free fermionic case the contribution from zero-mode $|\theta|(k=0,0)$ is subtle. From (\ref{albe_2}) one may find that it is because when $k=0$, $U_{-\infty}(k,t)$ and $V_{-\infty}(-k,t')$ are both zero vectors. A more profound reason might come from (\ref{bogo_t2}). When $k=0$, (\ref{bogo_t2}) turns into
\begin{equation}
\begin{split}
a_{-\infty}({0},t)=\alpha({0},t)a({0})-\beta({0},t)a^{\dagger}(0), \hfill \\
a_{-\infty}^{\dagger}(0,t)=\beta^*({0},t)a({0})+\alpha^*(0,t)a^{\dagger}(0). \hfill \\
\end{split}
\label{zeromode}
\end{equation}
Then anticommutator $a(0)^2 = (a^{\dagger}(0))^2=a_{-\infty}(0)^2 = (a_{-\infty}^{\dagger}(0))^2=0$ implies that 
\begin{equation}
\alpha(0,t) \beta(0,t) \equiv 0
\end{equation}
Combined with (\ref{al_be}), we find that $(|\alpha(0,t)|,|\beta(0,t)|) = (1,0)$ or $(0,1)$. Therefore the contribution from the zero-mode $|\theta|(0,t)=0, {\pi/2}$ \cite{myersfermion}. However, given that in the latter case (\ref{zeromode}) turns into 
\begin{equation}
\left. \begin{array}{*{20}{c}}
a_{-\infty}({0},t)=-e^{i \varphi(t)}a^{\dagger}(0), \hfill \\
a_{-\infty}^{\dagger}(0,t)=e^{-i \varphi(t)}a({0}). \hfill \\
\end{array} \right\} \Longrightarrow \beta(0,t) \equiv e^{i\varphi} =0,
\end{equation}
i.e. the zero-mode contribution $|\theta|(k=0,t)$ has to be zero in our case. The Bogoliubov transformation of the zero-mode is trivial, since Majorana fermions have an unpaired zero-mode.

As for nonzero-modes, Fig. \ref{f6} shows that single-mode contribution has a peak at some nonzero mode and it moves close to zero mode as quench becomes slower. Most of the contribution to complexity $\mathcal{C}^{(1)}(0)$ comes from modes that remain adiabatic, i.e. $|\theta|(k>k_c,0)$. According to (\ref{k_cond}), when $J\delta t$ increases, $k_c$ decreases, so that adiabaticity is moving toward $k =0$. On the other hand, when $t=0$, adiabatic contribution
\begin{equation}
|\theta|_{adia}(k,t=0) \equiv \frac{1}{2} \tan^{-1} \left| \frac{(m_{in}-m(t=0))G}{m_{in}\cdot m(t=0)+G^2} \right|
\end{equation}
monotonically decreases (as $k$ increases). However, $|\theta|(k,t) \to 0$ when $k \to 0$ (\ref{albe_2}). Thus there exists a peak somewhere around $k_c$. As $J \delta t $ increases, the peak moves toward $k=0$; more concretely, the peak of the exact $|\theta|$ moves along the curve of $|\theta|_{adia}$. 

Now since $k_c \to 0$ when $J\delta t \to \infty$, all of the contributions from non-zero modes are adiabatic, i.e. $|\theta|(k,0)=|\theta|_{adia}(k,0)$ for all $k \ne 0$. This implies that circuit complexity $\mathcal{C}^{(1)}(0)$ saturates to a constant value \[ \int_{0}^{\pi} {\text d}k|\theta|_{adia}(k,0)\] given that $|\theta|_{adia}(k,0)$ is $\delta t$-independent. For a large $\delta t$ that satisfies $J \delta t \gg 1/b$, the difference 
\begin{equation}
\begin{split}
\mathcal{C}^{(1)}(0) |_{\delta t} - \mathcal{C}^{(1)}(0)|_{\delta t \to \infty} =& \int_{0}^{k_c}{\text d}k \left[ |\theta|(k,0)- |\theta|_{adia}(k,0) \right] \hfill \\
 \le & \left[ |\theta|(k,0)- |\theta|_{adia}(k,0) \right]|_{k \to 0} \times k_c =-|\theta|_{adia}(k \to 0,0) \times k_c 
\end{split}
\end{equation}
since $|\theta|_{adia}(k,0)$ is monotonically decreases and $|\theta|(k,0)$ is non-negative. $|\theta|_{adia}(k \to 0,0)$ is $\delta t$-independent, and one can figure out that 
\[|\theta|_{adia}(k \to 0,0) \to \pi/2. \]
On the other hand, according to (\ref{k_cond}), when $\delta t$ is large, $k_c \sim \sin k_c \sim \delta t^{-1/2}$. Therefore, we can see
\begin{equation}
\mathcal{C}^{(1)}(0) |_{\delta t}=  \int_{0}^{\pi} {\text d}k|\theta|_{adia}(k,0) + \mathcal{O}(\delta t^{-1/2})
\end{equation}
This might explain why in slow quench regime ($J \delta t \gg 1/b$), $\mathcal{C}^{(1)}(0)$ in the fermionic theory saturates with the rate $\delta t^{-1/2}$. A more rigid argument is given in section \ref{ssq}.

\begin{figure}[H]
\centering
\subfigure[ECP-type-like]{ \label{f2a}
\includegraphics[width=0.45\textwidth]{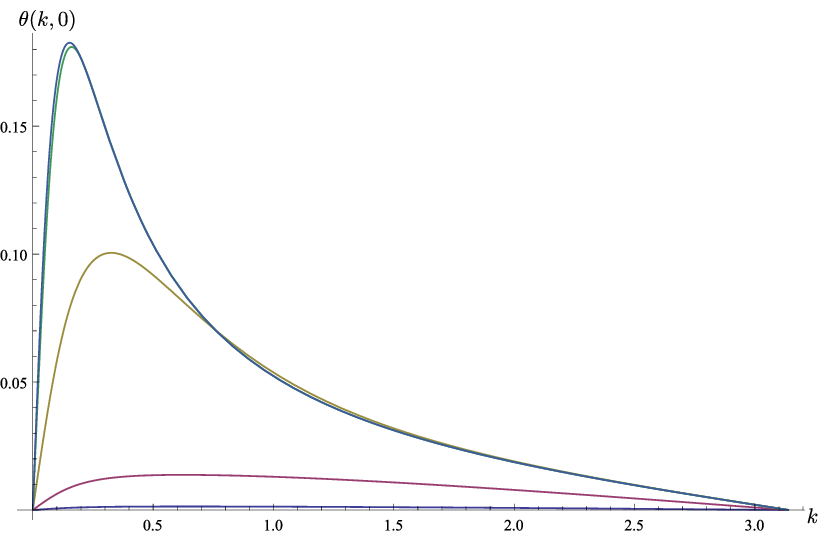}}
\subfigure[CCP-type-like]{ \label{f2b}
\includegraphics[width=0.45\textwidth]{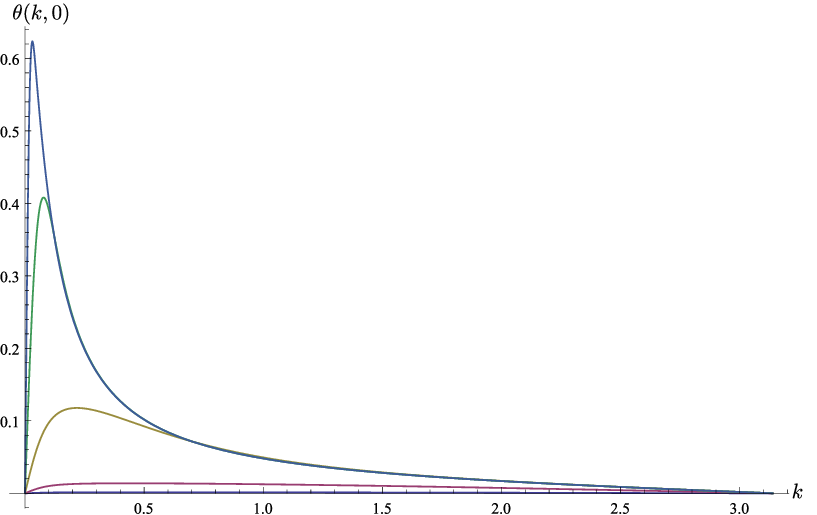}} \\
\caption{Single-mode contribution to complexity at $t=0$, $|\theta|(k,0)$ in ECP and CCP-like potentials when $b=0.01$. Purple, red, yellow, green and blue solid lines are $J\delta t = 0.01,0.1,1,10,100$, respectively.}
\label{f2}
\end{figure}

\begin{figure}[H]
\centering
\includegraphics[width=0.6\textwidth]{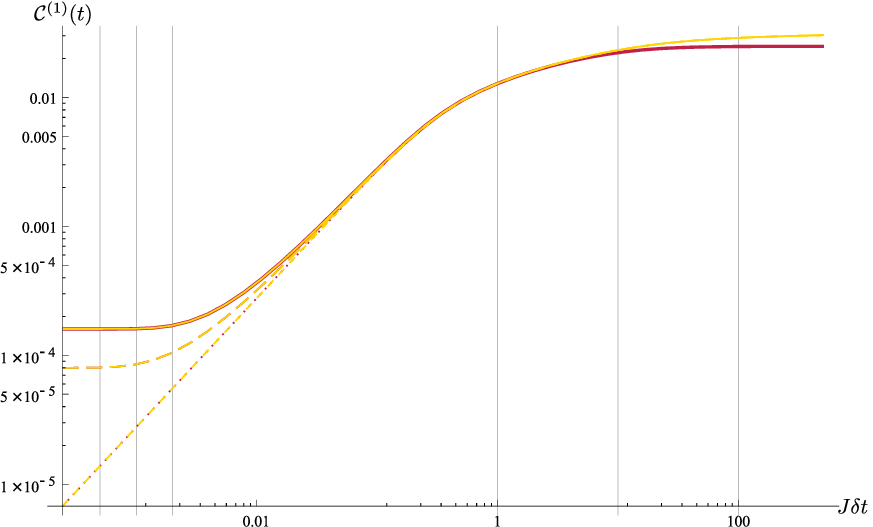}
\caption{Exact $\mathcal{C}^{(1)}(t)$-$J\delta t$ relations in log-log scale when $b=0.01$. Red and yellow lines correpond to ECP and CCP-type-like potential respectively. From solid to dashed, the curves correspond to $t=0.002, 0.001$ and $0.0005$, respectively. The plots differ at large $J\delta t$ (Red plots saturate more quickly).}
\label{f11}
\end{figure}

The theory with ECP-type-like potential is slightly different when slow quenched (Fig. \ref{f11}), complexity saturates much more quickly because the single-mode contribution saturates at large $J\delta t$ (Fig. \ref{f2a}).

\begin{figure}[H]
\centering
\subfigure[ECP-type-like: from solid to dotted lines $J\delta t=200,100,10,1,0.1,0.01$, respectively]{ \label{f1a}
\includegraphics[width=0.45\textwidth]{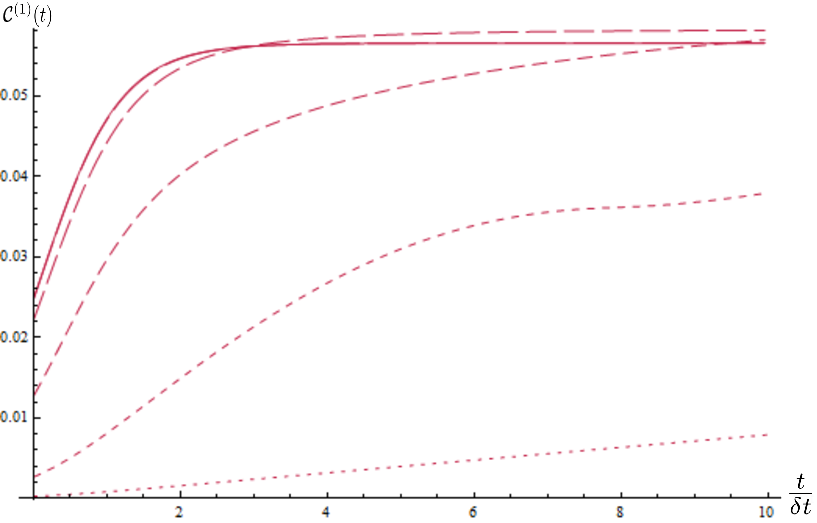}}
\subfigure[CCP-type-like: from solid to dotted lines $J\delta t=100,1,0.01$, respectively]{ \label{f1b}
\includegraphics[width=0.45\textwidth]{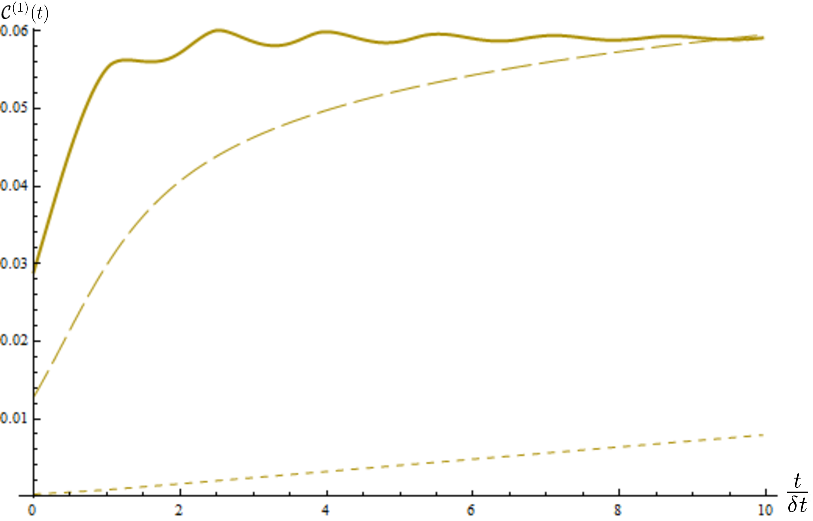}} \\
\subfigure[ECP-type-like: from solid to dashed lines $J\delta t=150,100$, respectively]{ \label{f1c}
\includegraphics[width=0.45\textwidth]{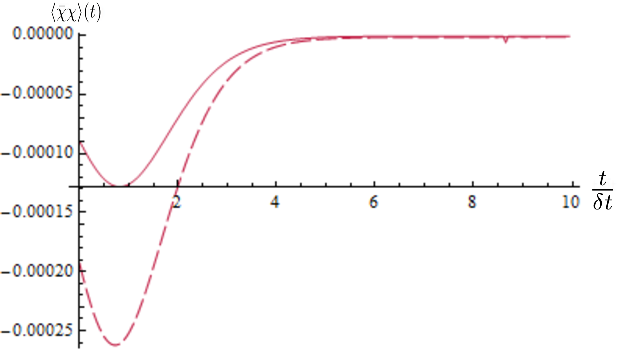}}
\subfigure[CCP-type-like: from solid to dashed lines $J\delta t=150,100$, respectively]{ \label{f1d}
\includegraphics[width=0.45\textwidth]{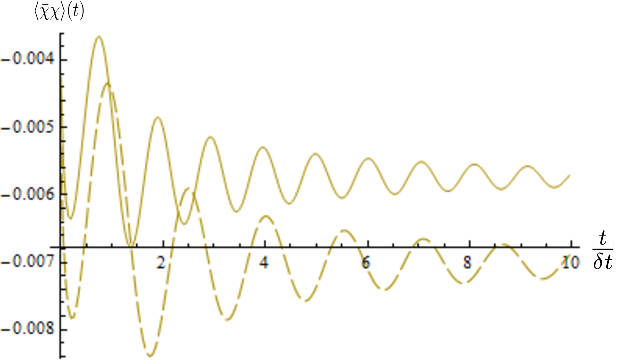}} \\
\caption{\ref{f1a}\&\ref{f1b}:Time evolution of complexity $\mathcal{C}^{(1)}(t)$ in ECP and CCP-like potentials; \ref{f1c}\&\ref{f1d}:Time evolution of $\langle \bar{\chi} \chi \rangle$ in ECP and CCP-like potentials. From thick solid lines to dotted lines $J\delta t$ decrease. Choose $b=0.01$.}
\label{f1}
\end{figure}

Finally, we numerically compare the late-time behaviors of circuit complexity (\cite{alves, jiang} have studied circuit complexity at late time in the bosonic free field with smooth quench and fermionic free field with instantaneous quench, respectively) in ECP-type-like potential ($a=1-b$) and CCP-type-like one ($a=1$), and the results are shown in Fig. \ref{f1}. It shows that circuit complexity of ECP-type saturates without oscillation, unlike CCP-type potential. This is consistent to quantities such as $\langle \bar{\chi} \chi \rangle$ (\cite{ddgms}). 

In conclusion, in this paper we have studied the scaling of circuit complexity in Majorana fermion field theory of 1D transverse field Ising model under quantum quench. It provides another evidence for the fact that just as correlation functions (e.g. $\langle \bar{\chi} \chi \rangle$) and entanglement entropy, complexity is a good quantity to see universal scaling in critical quench.

\section*{Acknowledgements}
I would like to thank Dr. Sumit R. Das, who suggested the problem and set it up. Without his help and valuable discussions, I would not be able to understand many issues about the problem and eventually complete this paper. The work is partially supported by the National Science Foundation grant NSF-PHY-1521045.

\end{document}